# Model for Predicting End User Web Page Response Time

Sathya Narayanan Nagarajan     Srijith Ravikumar

## 1. ABSTRACT

Perceived responsiveness of a web page is one of the most important and least understood metrics of web page design, and is critical for attracting and maintaining a large audience. Web pages can be designed to meet performance SLAs early in the product lifecycle if there is a way to predict the apparent responsiveness of a particular page layout. Response time of a web page is largely influenced by page layout and various network characteristics. Since the network characteristics vary widely from country to country, accurately modeling and predicting the perceived responsiveness of a web page from the end user's perspective has traditionally proven very difficult. We propose a model for predicting end user web page response time based on web page, network, browser download and browser rendering characteristics. We start by understanding the key parameters that affect perceived response time. We then model each of these parameters individually using experimental tests and statistical techniques. Finally, we demonstrate the effectiveness of this model by conducting an experimental study with Yahoo! web pages in two countries and compare it with 3$^{rd}$ party measurement application.

## 2. INTRODUCTION

The Internet has become an omnipresent part of everyday life of people around the world. Web response time is one of the key attributes for attracting a large user base as the web pages are served from remote location. According to various studies, Yahoo! witnessed a 5-9% drop in web traffic when web page response time was slower by 400ms, Amazon witnessed 1% fewer sales for every 100ms increase in response time (including content-heavy pages), Google witnessed 20% fewer searches when response time was increased by 500ms [2], Facebook witnessed 3% drop in traffic when pages were 500ms slower and 6% traffic drop when pages were 1000ms slower [3].

Response time of a web page predominantly depends on the number of HTTP objects, size of the objects, and the underlying throughput [4]. In order to be successful, web pages must strike the optimum balance between the content served and perceived end user response time.

### 2.1  Problem

For companies whose business is based largely (if not entirely) on internationally accessible web-based content -- like Yahoo! --  it is imperative to deliver the right content in a timely manner.  It is critical that the new web sites be designed from the start with end-user response time in mind; Slow or unresponsive web sites are likely to quickly lose potential customers out of frustration.  Additionally, addressing performance issues caused by poor page design after a site has been launched can be costly.

In order to build a faster performing, internationally aware web site, designers and developers need a model that can predict end user response time for a particular country during web page design and development phase.

### 2.2  Existing Approaches

The performance of Internet applications has been the topic of many scholarly research papers as well as industrial research. There have been various researches in the area of network performance modeling [6,7,8] , web server performance [9, 10,11,12] and web page response time modeling [13, 14, 15, 16]. While network performance model was more focused on TCP/IP modeling, web server performance modeling were based on probability and queuing theory more focused on response time of the server. Web page response time model focused on overall web page response time but each of these had limitations due to which it cannot be used for real life modeling

Thiam Kian Chew did a Ph.D. thesis on *Web Page Performance Analysis* (2009)[13] where he studied and modeled a web page response time.  However, Thiam Kian Chew's research ignores the effect of network latency & bandwidth, as can be inferred from this line in his thesis: *"Effects of Web server hardware and network related bandwidth, on Web page response time are not included as main parameters in the models."*  Based on other research work done in this area, it has been observed that more than 60% of the time is being spent on network for downloading HTTP components [4]. Since Thaim's research doesn't consider the network related characteristics in the model, it cannot be used in real time scenarios.

David P. Olshefski came up with a model called Certes [14,15] that provides server-side measure of mean client perceived response time that includes the impact of failed TCP connection attempts on Web server performance. But, certes model has ignored the effect of DNS lookups, rendering and the usage of Content Delivery Networks (CDN) [18]. Certes model needs measurement at server side which is not practically possible on CDN servers as CDN servers could belong to 3rd party service providers like Akamai. Since modern day web pages predominantly use CDNs, this model too cannot be used in real time scenarios.

Since there is no model available that can be used in real time, there was a need to do research in this field and build a predictive model based on web page complexity and network characteristics.

### 2.3 Our Approach
In this paper, we develop a new approach that considers web page characteristics, network characteristics, browser download and rendering characteristics for predicting the web page response time.

We obtain network characteristics for a particular country from Gomez[1] last mile data. Gomez is a 3rd party company, which measures web response time of an application on last mile of a particular country. We analyze more than half a million HTTP components response time to get the network related information.

For modeling browser characteristics, we use the vast expertise of our exceptional performance team's research [19] on web browser and perform various browser experiments.

Then, we use statistical techniques like curve fitting methodology to derive equations from the vast data obtained from various network and browser experiments.

### 2.4 Key Challenges
Our model faces two critical challenges. First**,** there are too many variants on network and there hasn't been any proven way to get network characteristics of a country. Second, modern day web browsers are getting optimized for web page performance in every release. They can download multiple HTTP components in parallel to optimize the I/O wait time. However, there is no way to measure/obtain the parallel download efficiency from Gomez or through any other tools.

### 2.5 Our Contributions
In this paper, we come up with a novel approach that overcomes the challenges mentioned above.

1) For modeling the network related parameters, we choose to use descriptive modeling techniques as opposed to constructive modeling techniques. Descriptive and Constructive modeling techniques are different modeling techniques that can be used to build models. If a model is derived from a set of measured data that summarizes the dataset by some statistical parameters, it is called a descriptive model. If a model is a description of a process or system elements that could have ideally produced the dataset to be examined, then it is called a constructive model [17]. We consider Network elements, Brower elements as a black box and use statistical techniques to build models.

We obtain network characteristics for a particular country from Gomez last mile data.

2) For browser characteristics, we use the vast expertise of our exceptional performance team's research to come up with a model for browser parallel download. We then use experiments to validate the accuracy of browser parallel download efficiency model

### 2.6 What this model doesn't cover
As much as we understand what this model is all about, we need to also understand its limitations.

- ✓ This model doesn't cover the server processing time. Modeling backend itself is a separate process and it depends on the underlying technical and deployment Architecture. Yahoo! web server performance is very much optimized and it takes less than 10% of the overall response time. Hence, based on the observation of multiple Y! web server processing time, we assume server processing time to be 200ms.**.**

✓ Model assumes base page to be served from Y! colo and static resources to be served from CDN. If this changes, then properties with similar deployment architecture need to be measured and modeled.

## 2.7 Organization of this paper

In the next section, we provide details on the components of web page response time and a mathematical equation for computing the response time. Section 4 explains modeling of various parameters involved. Section 5 provides the experimental result for Indonesia and US markets. Section 6 and 7 discuss the future work and our conclusion.

## 3. COMPONENTS OF A WEB PAGE RESPONSE TIME

At a high level, the perceived response time of a web page can be described as the time taken by the server for processing ($T_{sr}$), time taken for server-to-client transmission ($T_t$) and time taken by the browser for rendering ($T_r$)

$$T_{rt} = T_r + T_{sr} + T_t$$

$T_t$ can be further represented as time taken for DNS Lookup ($T_{dns}$), time taken for TCP/IP connection ($T_c$), time taken for first byte ($T_{fbt}$) and time taken for content download ($T_{cd}$).
$$T_t = T_{dns} + T_c + T_{fb} + T_{cd}$$

Since a web page is comprised of multiple HTTP objects, the transmission time can be further represented as:

$$T_{ti} = T_{dns} + T_c + \Sigma^n_{i=1} T_{fbi} + \Sigma^n_{i=1} T_{cdi}$$

Where $n$ is the number of HTTP objects, $T_{fbi}$ is the first byte time of $i$-th HTTP object and $T_{cdi}$ is the content download time of $i$-th HTTP object.

However, there are many additional factors such as the prevalence of content delivery networks (CDN), and parallel download capabilities of modern web browsers, which influence the transmission of HTTP objects. As such, a comprehensive web page performance model must include more than a simple summation of object transmission times.

### 3.1 Effect of Content Delivery Network

Modern day web architecture generally employs the use of CDNs; Only the base HTML is generated by the server, while static resources are served from CDNs, which are typically closer to the end user. Due to this factor, the time taken for base page needs to be computed separately as it involves server processing time and network latency that differs from that of the static components served by the CDN. Hence transmission time can be represented as sum of transmission time for base page ($T_{bp}$), transmission time taken for any additional static components ($T_{sc}$)

$$T_t = T_{bp} + \Sigma^n_{i=1} T_{sci}$$

$T_{bp}$ can be represented as sum of DNS Lookup time for base page ($T_{dnsbp}$), TCP/IP connection time for base page ($T_{cbp}$), first byte time for base page ($T_{fbbp}$) excluding the server time ($T_{sr}$) and content download time for base page ($T_{cdbp}$)

$$T_{bp} = T_{dnsbp} + T_{cbp} + T_{fbbp} + T_{cdbp}$$

$T_{sc}$ can be represented as the sum of DNS lookup time for the CDN ($T_{dnssc}$), TCP/IP connection time for CDN ($T_{csc}$), the sum of first byte time for each static resource ($T_{fbsci}$), and the sum of content download time for each static resource ($T_{cdsci}$)

$$T_{sc} = T_{dnssc} + T_{csc} + \Sigma^n_{i=1} T_{fbsci} + \Sigma^n_{i=1} T_{cdsci}$$

### 3.2 Effect of Browser Parallel Download

Modern web browsers are capable of downloading web objects in parallel, masking I/O wait time and effectively reducing perceived response time.

However, not all components can be downloaded in parallel. Browsers must comply with a few rules[5,19] regarding parallel downloads:

1. When downloading a page, the browser cannot download static resources described by the page's HTML until the HTML has been fully-downloaded and parsed, revealing the address of the related static resources.

2. When a browser encounters a javascript function, it doesn't do parallel download as it cannot determine whether any execution of JS is needed for further processing.

3. When one of the last few components is 25+KB more than the other components, then there won't be any effect of parallel download for this component, as the content of this component will be still getting downloaded while other components would have finished downloading.

Based on these rules, the transmission time can be represented as:

$$T_t = T_{bp} + (\Sigma^m_{j=1} T_{dnssci} + \Sigma^n_{j=1} T_{csc\,j})/BPE + \Sigma^o_{k=1} T_{jsk} + ((\Sigma^p_{l=1} (T_{otl}))/BPE)$$

Where $T_{dnssc}$ is the time taken for DNS lookup for a particular domain, $m$ is the number of unique domains available in the web page. $T_{csci}$ is the TCP/IP connect time for $i$-th element, $n$ is the number of parallel connections. $T_{jsk}$ is the transmission time for java script component for the $k$-th element, $o$ is the number of JS components. $T_{otl}$ is the transmission time for the $l$-th element of static resource other than JS, $p$ is the total number of static resource other than JS. BPE is Browser Parallel Efficiency.

$T_{js}$ can be further represented as $T_{jsfb} + T_{jscd}$ where $T_{jsfb}$ is the first byte time taken for js and $T_{jscd}$ is the content download time for java script:

$$T_{js} = T_{jsfb} + T_{jscd}$$

$T_{ot}$ can be further represented as $T_{otfb} + T_{otcd}$, where $T_{otfb}$ is the first byte time taken for static resources excluding Javascript and $T_{otcd}$ is the content download time for static resources excluding Javascript:

$$T_{ot} = T_{otfb} + T_{otcd}$$

Updated transmission time

$$T_t = T_{dnsbp} + T_{cbp} + T_{fbbp} + T_{cdbp} + (\Sigma^m_{j=1} T_{dnssci} + \Sigma^n_{j=1} T_{csc\,j})/BPE) + \Sigma^o_{k=1} (T_{jsfbk} + T_{jscdk}) + ((\Sigma^p_{l=1} (T_{otfbl} + T_{otcdl}))/BPE)$$

Based on this equation, following tables (Table 1 and Table 2) provides variables that need to be modeled for web page response time.

| Notation | Description |
|---|---|
| $T_{dnsbp}$ | Base Page DNS Lookup Time |
| $T_{cbp}$ | Base Page Connect Time |
| $T_{fbbp}$ | Base Page First Byte Time |
| $T_{cdbp}$ | Base Page Content Download Time |
| $T_{dnssc}$ | Static Component DNS Lookup Time |
| $T_{csc}$ | Static Component Connect Time |
| $T_{jsfbc}$ | Java Script First Byte Time |
| $T_{jscd}$ | Java Script Content DL Time |
| $T_{otfb}$ | Other Static components First Byte Time |
| $T_{otcd}$ | Other Static components Content DL Time |

| Notation | Description |
|---|---|
| BPE | Brower Parallel Download Efficiency |
| Tr | Browser Rendering time |

*Table 1: Network Variables*

*Table 2: Browser Variables*

## 4. MODELING WEB PAGE RESPONSE TIME

This section explains the experimental setup and approach taken for modeling the parameters involved in web page response time.

### 4.1 Experimental Setup

We devised multiple approaches to get predictable values for each of the parameters that needed to be modeled. Following approaches were used to conduct the research

**Network parameters:** We used Gomez last mile data for measuring the network related information. Gomez last mile uses machines from various ISPs in a particular country. It provides DNS time, connect time, first byte time and content download time of each component. This gives a widespread data across the country that can be used to model the network parameters. We ran last mile tests for multiple Yahoo! web pages and collected these network related information for half a million HTTP components.

**Browser Parameters:** Browser related parameters like Parallel Download and Rendering were tested using Firefox browser. Various tools like YSlow, Firebug, NetExport were used for obtaining this information

**Statistical Model:** Results got from Gomez last mile tests and browser experiments were analyzed using statistical models like curve fitting methodology.

## 4.2 Model

**Base Page DNS Lookup Time($T_{dnsbp}$)**

We derive base page DNS lookup time from the DNS lookup time of various Y! pages DNS time. Since, all Y! pages use a shared name server, it is similar for all these pages. Hence, we use average value of DNS time obtained from various Y! pages. Following table provides the DNS response time of ID News, ID OMG Homepage, ID News Sports, ID Search and ID Frontpage were analyzed and the mean of the DNS time has been taken as Base Page DNS Lookup Time($T_{dnsbp}$) for ID. Following table represents the ID properties and their Average DNS values

| Property | DNS Time (ms) |
|---|---|
| ID News | 169.23 |
| ID OMG Homepage | 146.84 |
| ID News Sports | 170.33 |
| ID SRP | 176.67 |
| ID Frontpage | 149.70 |
| **Mean** | **162.55** |

*Table 3: Base Page DNS Lookup Time*

**Base Page Connect Time ($T_{cbp}$)**

Similar to $T_{dnsbp}$, base page connect time can be derived from average of connect time of the Y! Properties measured from last mile. Since all Y! Properties for Indonesia are predominantly served from Singapore, connect time for these properties will be similar. In this study, base page connect time ($T_{cbp}$) has been derived from the average connect time of ID News, ID OMG Homepage, ID News Sports, ID Search and ID Frontpage. Following table represents the ID properties and their average Connect time .

| Property | Connect Time (ms) |
|---|---|
| ID News | 193.13 |
| ID OMG  Homepage | 233.72 |
| ID News Sports | 135.43 |
| ID SRP | 130.37 |
| ID Frontpage | 139.77 |
| **Mean** | **166.484** |

*Table 4: Base Page TCP/IP Connect Time*

**Base Page Content Download Time ($T_{cdbp}$)**

Time taken for downloading content from a remote location is directly proportional to the size of the data. If the file size downloaded is small, time taken will be less and vice versa. In order to derive the equation between the file size and content download time taken at Indonesia, base page of the properties were plotted on a XY axis with X-axis representing the base page size and the Y-axis representing content download time taken. The resultant graph was a linear curve as represented below.

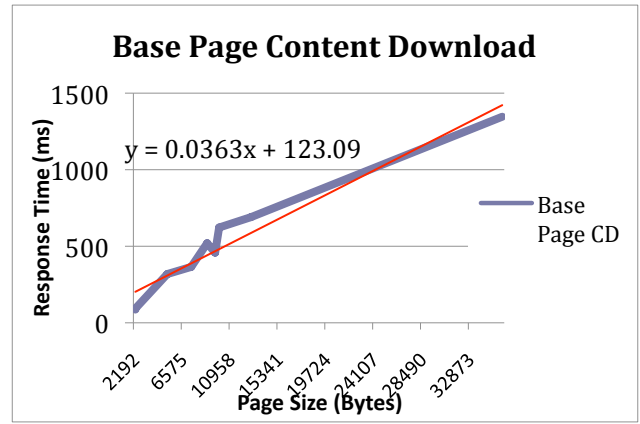

*Figure 1: Base Page Content Download Time*

**Base Page First Byte Time ($T_{fbbp}$)**

Base Page First Byte Time comprises of network latency and server processing time. Since server-processing time varies from property to property, it is assumed to be a constant in this model. Network latency is agnostic of any property. Hence, we derive First Byte time for Indonesia using the following way

- Average First Byte time (FB) of various Y! properties were measured from last mile Gomez test.

- Average server processing time (SP) of Y! properties were measured through Argus.

- The difference between First Byte time and Server processing time provides the perceived network latency for that property.

$$T_{fbbpi} = FB_i - SP_i$$

Where i is the i-th property. $T_{fbbpi}$ provides Base Page First Byte Time for i-th property.

Base Page First Byte Time for a country can be derived from taking the mean of all the properties:

$$T_{fbbp} = (\Sigma^{n}_{i=1} T_{fbbpi})/ i$$

Following table provides the computed base page first byte time:

| Property | First Byte Time (ms) | Argus Server Time (ms) | Response Time (ms) |
|---|---|---|---|
| ID Frontpage | 594.539375 | 210 | 384.54 |
| ID News | 762.8876463 | 140 | 622.89 |
| ID News Entertainment | 542.1180042 | 50 | 492.12 |
| ID OMG Homepage | 691.0155561 | 320 | 371.02 |
| ID News Technology | 523.4878741 | 140 | 383.49 |
| | | Mean | 450.81 |

*Table 5: Base Page First Byte Time*

**Static component DNS Lookup Time** ($T_{dnssc}$)
Most of the Yahoo! web pages use Yahoo! Caching Service (YCS) for static components. We measure the DNS time for YCS domain (l.yimg) using Gomez last mile test and obtain the DNS lookup time. Average DNS lookup time for l.yimg in Indonesia is **148ms**

**Static component Connect Time** ($T_{csc}$)

We measure the connect time for l.yimg urls using Gomez last mile test and obtain the connect time for the CDN. Average connect time for l.yimg in Indonesia is **163ms**.

**Static Component First Byte Time ($T_{fbsc}$)**

Y! web pages use content delivery network for downloading the static components like CSS, Javascript and Image. Since, static components don't require any server processing, first byte time for static components are resultant of the network latency between the end user and CDN. First Byte Time of static component can be modeled by studying the influence of page size on the First Byte time. To understand this, Average first byte time of HTTP components varying from 1KB to 80KB from Gomez Last Mile were plotted on a XY axis. X axis represented the Page Size and Y Axis represented the first byte time. By using curve fitting technique, a line was plotted to derive the equation. Following graph represents the static component first byte time model for Indonesia.

*Figure 2: Static Component First Byte Time*

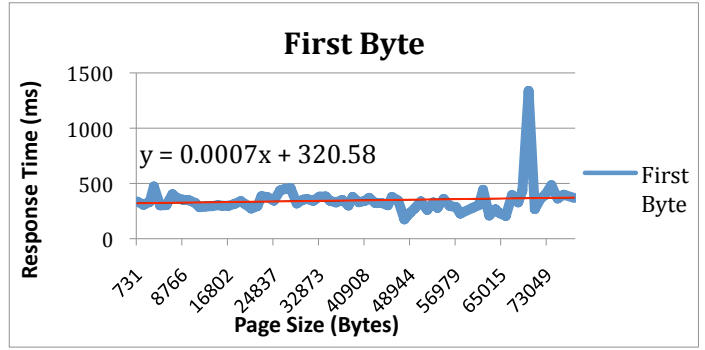

**Static Component Download Time ($T_{cdsc}$)**
Similar to the content download time of base Page, content download time of static component is directly proportional to the size of the component. To find the correlation between the page size and time taken, HTTP components ranging from 1KB to 80KB files were plotted on a XY axis. X axis represented the page size and Y axis represented the content download time. By using curve-fitting technique, a line was plotted to derive the equation. Following graph represents the static components content download time model for Indonesia

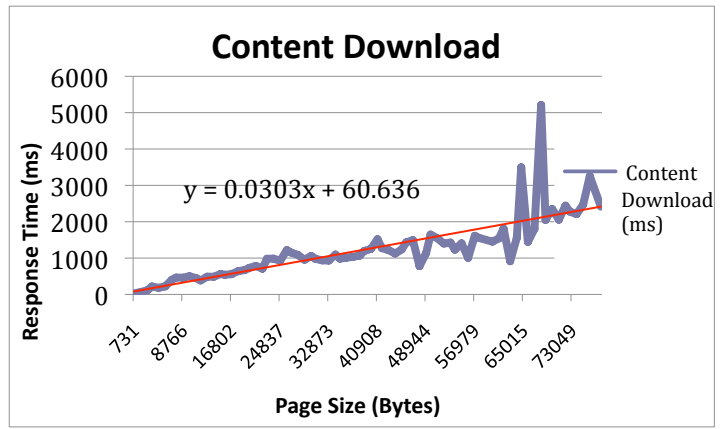

*Figure 3: Static Component Download Time*

## Browser - Parallel Download Efficiency (BPE)

Modern day web browsers download multiple HTTP components in parallel to improve the overall response time of a page. Let us illustrate this with an example:

Following figure represents the waterfall chart of 4 HTTP components being downloaded 1 at a time. Total response time for all 4 components to be downloaded is 4 seconds.

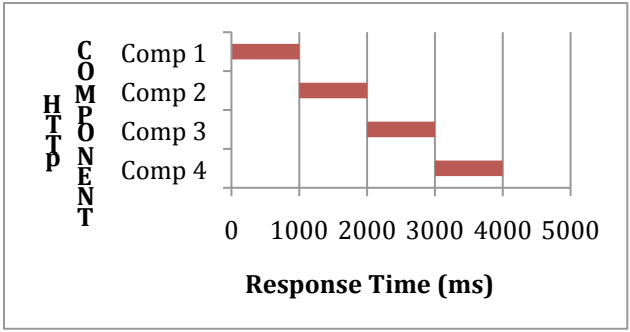

**Figure 4: HTTP component downloaded 1 at a time**

Following figure represents the waterfall chart of 4 HTTP components being downloaded 2 at a time.

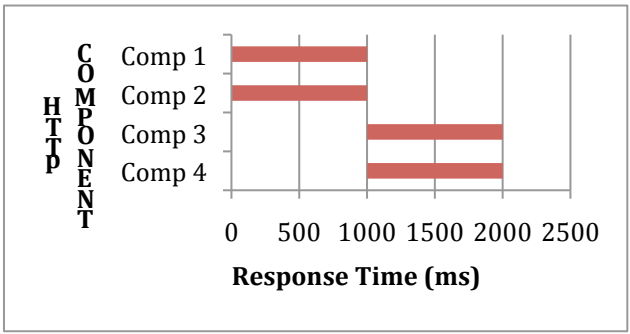

**Figure 5: HTTP component downloaded 2 at a time**

In theory, if browser downloads 2 at time, total response time can be 2 seconds.

Following figure represents the waterfall chart of 4 HTTP components, where all the components are being downloaded in parallel.

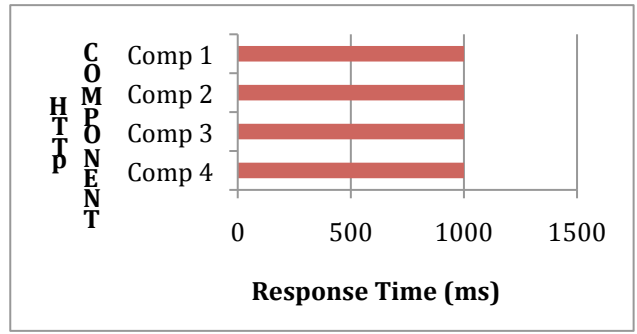

**Figure 6: HTTP components downloaded 4 at a time**

In theory, if all 4 components are downloaded in parallel, total response time can be 1 second.

The reason for response time improvement is due to the efficiency in network utilization. While the server is waiting for the response for the first request, it can use the network to download other resources. But, this depends entirely on the availability of network bandwidth. If there are 2 parallel requests and the network bandwidth utilization is already at maximum, then downloading one more request will not improve the response time, as the content download time will be high. Hence, when the browser makes a request for one HTTP component, it can only use the first byte time to download other HTTP components in parallel. Let us take an example where there are 5 HTTP components with 100ms content download time each. Assuming that the first byte time for one of the component is 400ms, browser can download other 4 components in this 400ms delay. Parallel download efficiency for this is 5.

The number of parallel download depends on the first byte time and the average content download time. The efficiency of the parallel download capability can be derived using this equation

BPE = Browser request + ( First Byte time / Avg content download time taken for HTTP Component)

In order to validate this equation, we conducted a test varying the number of browser download connections on various Y! Properties. This test was done on a broadband connection in India.

### Test Setup details

| Network Bandwidth(kbps) | 1319 |
|---|---|
| First Byte Time (ms) | 82.59 |

*Table 6: Test setup details*

### IN Frontpage
For, IN Frontpage, we derived browser parallel efficiency as 3 through our equation and this was validated with the best response time got from the experiment

#### By Equation

| Property | Page Size | Total Static Request | Avg Page Size | Avg content Download time (ms) | BPE |
|---|---|---|---|---|---|
| In Frontpage | 304.8 | 38 | 64.17 | 48.65 | 3 |

*Table 7: IN Frontpage Browser Equation*

#### By Experiment

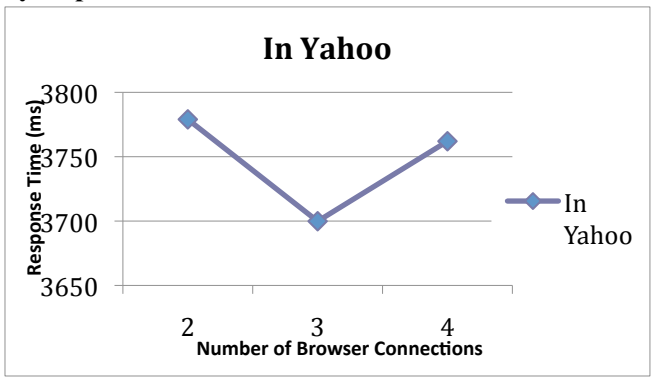

*Figure 7: IN Frontpage browser experiment*

### IN OMG
For, IN OMG, we derived browser parallel efficiency as 2 through our equation and this was validated with the best response time got from the experiment

#### By Equation

| Property | Page Size | Total Static Request | Avg Page Size | Avg content Download time (ms) | BPE |
|---|---|---|---|---|---|
| IN OMG | 576.6 | 32 | 144.15 | 109.29 | 2 |

*Table 8: IN OMG Browser Equation*

#### By Experiment

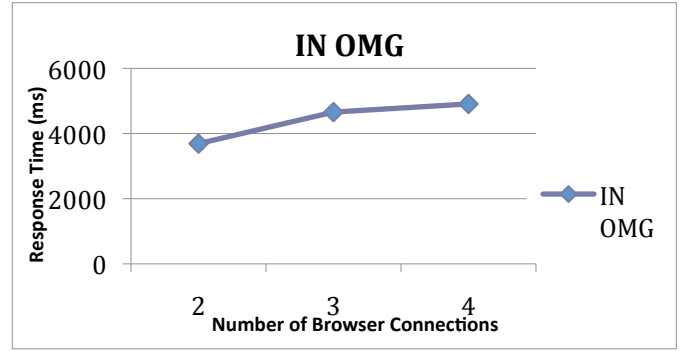

*Figure 8: IN OMG browser experiment*

### Browser rendering (Tr)

The final part is to model the rendering of the page. While browser can do this in parallel to the download of HTTP components, still significant amount of time is spent in rendering the Web page. This can be noticed in the net panel of Firefox where there are instances of no activity for 100-200ms. Hence it is imperative to know the time taken for rendering a web page to get a holistic view.

There isn't any straightforward way to get these values through a plug-in. 25+ Y! Web pages of varying page weight were saved locally and time taken for rendering was measured. Saving the web page locally removes the network latency and it does what the browser does after downloading the components.

Based on the observation of various Y! pages, we chose to use average KB per HTTP request as the measure of complexity. While there are multiple components like CSS, JS, Flash, IFrames that can influence the rendering, one of the key process of rendering is to fetch the components from the file system before rendering. Since this involves I/O operation, it was taken as the key attribute for rendering time for our model. Based on this attribute, web pages can be classified into 3 different categories Simple, Medium and Complex.

| Page Type | N = Avg KB / HTTP Req | Equation |
|---|---|---|
| Simple | N >16 | y = 0.0008x - 0.0271 |
| Medium | 11> N <16 | y = 0.4323ln(x) - 2.0771 |
| Complex | N <11 | y = 0.55ln(x) - 2.6079 |

*Table 9: Browser Rendering*

Where N = Avg KB per HTTP Request

## 5. EXPERIMENTAL RESULTS

For validating the results, an emerging market and a developed market were chosen to ensure that the model could work in different conditions.

Following tables provide comparison of response time obtained from model with Gomez Last Mile test

**Indonesia:**

| Property | Predicted (ms) | Gomez (ms) |
|---|---|---|
| US FP | 3237.89 | 3084 |
| US News | 4562.72 | 4894 |
| US OMG Homepage | 3234.63 | 3007 |
| US News Entertainment | 3142.53 | 3348 |
| US OMG Celebs | 3078.00 | 3248 |

**Table 10: ID Experimental Results**

*For Indonesia model can predict within 2.08% of Gomez value with standard deviation of 1.36%*

**USA:**

| Property | Predicted (ms) | Gomez (ms) |
|---|---|---|
| ID FP | 10260.43 | 10567 |
| ID OMG | 15504.78 | 15154 |
| ID News | 10301.52 | 10213 |
| ID Ent Article | 6156.92 | 6147 |
| ID OMG Fotos | 14051.69 | 14386 |
| ID OMG News All | 15969.20 | 16615 |

**Table 11: US Experimental results**

*For USA, model can predict within 6.53% of Gomez value with standard deviation of 1.98%*

For more details on the application of model, sample worksheet for ID OMG has been given as part of Appendix 1

## 6. FUTURE WORK
- ✓ Model can be made more accurate by collecting more Last Mile data and further tuning the equations.
- ✓ Currently rendering model is based on Firefox. Rendering model of IE and Chrome can be derived so that complexity of rendering can include the % of browser market share.
- ✓ Further research can be done to identify and adopt any other industry research to improve the model's process and improve the efficiency.

## 7. CONCLUSION

Performance of an application is essential to attract a large user base. Having a predictive model for end user response time aids in building high performing web sites right the first time. Due to varied complexity, there has never been a model to predict end user response time. This paper has provided a way to model the end user response time and experimentally proved that it can predict response time comparable to that of actual value obtained through Gomez.

## 9. APPENDIX – 1

**Response Time computation for ID OMG**

| Sl No | HTTP Component | Type | Size (Bytes) | CD (ms) | FB (ms) | SUM (ms) |
|---|---|---|---|---|---|---|
| 1 | http://id.omg.yahoo.com/ (C0) | text/html | 18799 | 661.36 | 498 | 1159.36 |
| 2 | /d/combo(C1) | text/css | 10013 | 364.03 | 327.5091 | 337.34 |
| 3 | /jj/image-4cfca6f7bd139-298cr_zaskia_irwansyah.jpg(C2) | image/jpeg | 14255 | 492.56 | 330.4785 | 401.48 |
| 4 | /jj/image-4cfca6ee30633-298_bachdim.jpg(C1) | image/jpeg | 23753 | 780.35 | 337.1271 | 545.11 |
| 5 | /jj/image-4cfca89fbec3c-298omg_olla.jpg(C2) | image/jpeg | 14514 | 500.41 | 330.6598 | 405.40 |
| 6 | /jj/image-4ce403be14ed3-298_william_middleton.jpg(C2) | image/jpeg | 18439 | 619.34 | 333.4073 | 464.75 |
| 7 | /jj/image-4c451fa497e71-hugh_jackman_298.jpg(C1) | image/jpeg | 14101 | 487.90 | 330.3707 | 399.15 |
| 8 | /jj/image-4c0e095faf912-298okezone_dewipersik.jpg(C2) | image/jpeg | 14940 | 513.32 | 330.958 | 411.84 |
| 9 | /jj/image-4cfca6f219c00-298_rapunzel.jpg(C1) | image/jpeg | 13781 | 478.20 | 330.1467 | 394.32 |
| 10 | /jj/image-4c7b68f24a37f-george+clooney+298+wire.jpg(C2) | image/jpeg | 17023 | 576.43 | 332.4161 | 443.34 |
| 11 | /jj/image-4cfb85a948ef3-298_wsatcc.jpg(C1) | image/jpeg | 19916 | 664.09 | 334.4412 | 487.09 |
| 12 | /jj/image-4cfb86066c0d0-298igt_waysy.jpg(C2) | image/jpeg | 15666 | 535.32 | 331.4662 | 422.82 |
| 13 | /jj/image-4cfca6e8d94e4-298viva_ariel.jpg(C1) | image/jpeg | 12462 | 438.23 | 329.2234 | 374.37 |
| 14 | /jj/image-4cac151b16d9a-298_mayer.jpg(C2) | image/jpeg | 15281 | 523.65 | 331.1967 | 417.00 |
| 15 | /jj/image-4c85adfa20648-miley+cyrus+298+wireimages.jpg(C1) | image/jpeg | 14661 | 504.86 | 330.7627 | 407.62 |
| 16 | /d/combo(C2) | javascript | 68373 | 2132.34 | 368.3611 | 2500.70 |
| 17 | /d/i/id/omg/bg_page_tile.png(C1) | image/png | 159 | 65.45 | 320.6113 | 188.32 |
| 18 | /d/i/id/omg/bg_page.png(C1) | image/png | 10794 | 387.69 | 328.0558 | 349.15 |
| 19 | /d/i/id/omg/dropshadow_bg.png(C1) | image/png | 231 | 67.64 | 320.6617 | 189.41 |
| 20 | /d/i/id/omg/sprite_header2.png(C1) | image/png | 77188 | 2399.43 | 374.5316 | 1353.15 |
| 21 | /d/i/id/omg/sprite_repeat5.png(C2) | image/png | 628 | 79.66 | 320.9396 | 195.42 |
| 22 | /d/i/id/omg/sprite_all5.png(C2) | image/png | 26972 | 877.89 | 339.3804 | 593.79 |
| 23 | /combo(C3) | x-javascript | 27212 | 885.16 | 339.5484 | 1224.71 |
| 24 | s=21429..._pn%032142987015%04A_%031 (C4) | image/gif | 43 | 61.94 | 320.5301 | 186.57 |
| 25 | /p.pl(C5) | image/gif | 0 | 60.64 | 320.5 | 185.92 |
| 26 | /itr.pl(C6) | image/gif | 43 | 61.94 | 320.5301 | 186.57 |
| 27 | /searchassist(C0) | application/json | 6334 | 373.42 | 498 | 425.08 |
| | BROWSER PARALLEL EFFICIENCY | | 2.05 | | | |
| | $T_{dnsbp}$ | | 182 | | | |
| | $T_{cbp}$ | | 162 | | | |
| | $T_{fbbp} + T_{cdbp}$ | | 1159.36 | | | |
| | $T_{dnssc}$ | | 148 | | | |
| | $T_{csc}$ | | 163 | | | |
| | $\Sigma^{26}_{i=1} T_{fbsc} + T_{cdsc}$ | | 13490.43 | | | |
| | $T_{sr}$ | | 200 | | | |
| | **Total Response Time (predicted)** | | **15504.78** | | | |
| | **Gomez Response Time (Actual)** | | **15154** | | | |
| | **Predicted vs Actual** | | **2.31%** | | | |